\documentclass[journal=jpcbfk,manuscript=article]{achemso}
\setkeys{acs}{usetitle = true}
\usepackage{color}
\usepackage{amsmath,amssymb}

\newcommand{\Q}{\mathbb Q}
\newcommand{\N}{\mathbb{N}}

\newcommand{\M}{\+M}

\author{Pablo Turjanski}
\email{pturjanski@dc.uba.ar}
\phone{ (++54 +11) 4576-3301 }
\affiliation[UBA-CONICET-ICC]{Departamento de Computaci\'on, Facultad de Ciencias Exactas y Naturales, UBA-CONICET-ICC, Buenos Aires, Argentina.}

\author{Diego U. Ferreiro}
\email{ferreiro@qb.fcen.uba.ar}
\phone{ (++54 +11) 4576-3300 }
\affiliation[UBA-CONICET-IQUIBICEN]{Protein Physiology Lab, Departamento de Qu\'imica Biol\'ogica, Facultad de Ciencias Exactas y Naturales, UBA-CONICET-IQUIBICEN, Buenos Aires, Argentina.}

\title{On the Natural Structure of Amino Acid Patterns in Families of Protein Sequences}

\begin{document}
\begin{abstract}

All known terrestrial proteins are coded as continuous strings of $\approx$20 amino acids. The patterns formed by the repetitions of elements in groups of finite sequences describes the natural architectures of protein families. We present a method to search for patterns and groupings of patterns in protein sequences using a mathematically precise definition for 'repetition', an efficient algorithmic implementation and a robust scoring system with no adjustable parameters. We show that the sequence patterns can be well-separated into disjoint classes according to their recurrence in nested structures. The statistics of pattern occurrences indicate that short repetitions are enough to account for the differences between natural families and randomized groups by more than 10 standard deviations, while patterns shorter than 5 residues are effectively random. A small subset of patterns is sufficient to account for a robust ''familiarity'' definition of arbitrary sets of sequences.

\end{abstract}

Keywords: protein sequence ; protein family ; patterns ; 


\small

\section*{Introduction}
\textit{``See first, think later, then test. But always see first. Otherwise you will only see what you were expecting.'' Douglas Adams} 

Protein molecules can be described as finite linear strings of $\approx$ 20 amino acid types. It is still an intriguing fact that most natural amino acid strings appear to be close to random by many statistical tests \cite{WEISS2000379}, yet most of the random polypeptides synthesized do not behave as proteins, as they do not fold to specific structures nor \textit{function} in a cellular context. Thus, the reduction in the description of proteins to linear strings of single amino acids misses a fundamental aspect to account for the, admittedly complex, biophysics of protein folding and function \cite{Wolynes:2012fk, Eaton:2017uq}. The search for 'structural codes' in the analysis of protein sequences must consider the occurrence of correlations in the patterns of groups of amino acids, a task that gets combinatorially prohibitive to analyze exhaustively for all protein sequences \cite{Dryden:2008kx}. Multiple heuristics designed to analyze correlations of amino acid patterns led to useful ways for approximating the grouping of sequences into families \cite{Doolittle:2010vn}, the mean structural ensembles of folded globules \cite{Morcos:2011zr}, and even hint to the thermodynamic folding behavior \cite{Schafer:2014cr} and evolutionary history of natural proteins \cite{Morcos:2014ys}. Most of these methods require multiple sequences to be aligned to a common matrix in a so-called multiple sequence alignment, a mathematical problem that is still open and thus the current computations need to be tediously curated by human experts \cite{Dickson:2010nx}.

To search for patterns and groupings of patterns in protein sequences, we use a mathematically rigorous definition of repetition and develop a characterization method with no adjustable parameters. A maximal repetition (MR) is a well-defined exact match of a continuous block of amino acids that occurs two or more times in one or more proteins, and any of its extensions (to the N-terminus, the C-terminus, or both) occurs fewer times. The search of MR can be implemented with an algorithm whose computational complexity is $O(n \log \, n)$, where $n$ is the size of the amino acid dataset, allowing for a very efficient exhaustive search \cite{Becher2009}. The natural architectures of protein sequences can be analyzed by the occurrence of MR patterns. In previous work \cite{Turjanski2016} we introduced the concept and defined a continuos $familiarity$ function that provides a fast quantification of the likelihood of any amino acid string to belong to a given set of sequences. This $familiarity$ function is computed from the search and match of MRs in sets of sequences. Here we show that the total MR set can be well separated into disjoint classes according to their recurrence in nested structures. We analyze the statistics of MR classes in several natural protein families and random strings, and find that only a small subset of MRs is sufficient to account for a robust ''familiarity'' definition. 

\section*{Methods}

\subsection*{Notation and Definitions}

Let there be an alphabet $\Sigma$, a finite set of symbols. We will consider linear sequences $s$ of symbols in $\Sigma$ of length $|s|$. We address the positions of a sequence $s$ by counting from 1 to $|s|$. A string $s[i..j]$ denotes the sequence that starts in position $i$ and ends in position $j$ in $s$. If  $1 \le i \leq j \leq |s|$ is false, then $s[i..j]$ is equal to the empty sequence. We say $u$ occurs in $s$ if $u=s[i..j]$ for some $i$, $j$. 
A right extension of an occurrence $u=s[i..j]$ exists if $j<|s|$ and is $s[i..j+1]$.
A left extension  of an occurrence $u=s[i..j]$ exists if $i>1$ and is $s[i-1..j]$.
A right context of an occurrence $u=s[i..j]$ exists if $j<|s|$ and is $s[j+1]$.
A left context  of an occurrence $u=s[i..j]$ exists if $i>1$ and is $s[i-1]$.

\textbf{Definition 1.} (Gusfield\cite{Gusfield}) A maximal repetition (MR) is a sequence that occurs more than once in $s$, and each of its extensions occurs fewer times. 

We classify the different MR patterns in three disjoint subsets: a) Super Maximal Repetition (SMR): a sequence that occurs more than once in $s$, and each of its extensions occurs only once; b) Nested Maximal Repetition (NE): a MR where all of its occurrences are contained in a longer MR; and c) Non-Nested Maximal Repetition (NN): a MR where at least one of its occurrences are not contained in a longer MR and is not SMR. Formal details on these definitions are described in previous work \cite{Taillefer} . 

\begin{figure}
 \center
 \includegraphics[scale=.4]{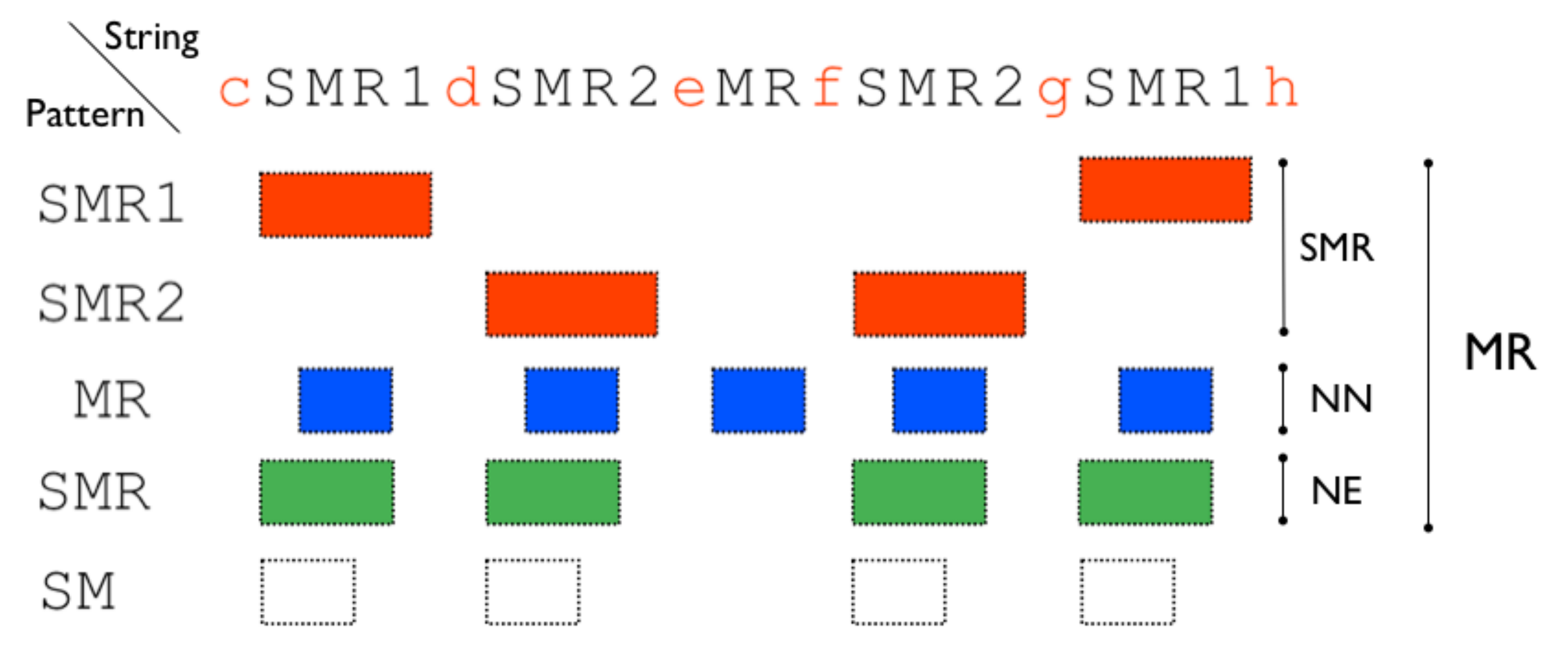}
 \caption{Maximal Repetitions (MR) computed for the input string shown on top (non-repeating symbols are in red lower case). MR patterns are classified in disjoint subgroups according to their patterns of occurrence as Super Maximal Repetition (SMR), Non-Nested Maximal Repetition (NN) and Nested Maximal Repetition (NE).} 
 \label{fig:MER_Example}
\end{figure}

An illustration of the proposed classification of MRs is presented in Fig. \ref{fig:MER_Example}. 
The set of MRs of the string $s1=cSMR1dSMR2eMRfSMR2gSMR1h$ (non-repeating symbols are lower case) is $\{MR, SMR, SMR1, SMR2\}$  (see Fig. \ref{fig:MER_Example}). Observe that $SMR1$ and $SMR2$ substrings are the longest MRs, occurring twice each. $SMR$ and $MR$ substrings are also MR because they occurs four and five times in $s1$ respectively, and each of their extensions occurs fewer times. Note that $SM$ is not a MR because $SMR$ (which is its unique possible repetitive right-extension) occur four times, violating the definition that any extension must occur fewer times. $SMR1$ and $SMR2$ are SMR because their extensions occur only once. $SMR$ is a NE since all of it occurrences are contained in $SMR1$ and $SMR2$. Finally, $MR$ is classified as NN since, although 4 of its occurrences are contained in longer patterns ($SMR1$ and $SMR2$), there is a fifth occurrence that is not contained in any other longer repetition.

\textbf{Definition 2.} Let $S$ be a set of $n$ sequences over the alphabet $\Sigma$, $S=\{s_1, s_2, ... s_n\}$. The set of MR in $S$ is the MR set of the sequence obtained by concatenation of all sequences in $S$, interleaved with different symbols $\$_1, ..., \$_{n-1}$ that are not in $\Sigma$. Thus, the set of MR in $S$ is the set of MR in $s_1\$_1s_2\$_2 ... \$_{n-1}s_n$.
If we work with sequences of characters, at the time of implementing this solution in a digital computer, there is an upper limit given by the necessary finite alphabet ($\Sigma$ has only 256 symbols in the extended ASCII table) which restricts the amount of different $\$$ symbols we can use, and thus the number of sequences we can concatenate. To overcome this limitation, we implement a logically distinguished symbol ($+$), which we assign to it the unique property of being different from itself. That is, $+\neq +$. Taillefer \textit{et. al.}\cite{Taillefer} proposed an algorithm that efficiently identifies and classifies MR from a sequence $s$ into SMR, NE and NN. We extended this algorithm in order to identify and classify MR originating from an arbitrary set of sequences (see Supplementary Methods section). 

\section*{Results and Discussion}

\subsection{Occurrences of MR patterns in natural proteins}

\begin{figure}[h]
 \center
  \includegraphics[width=0.9\textwidth]{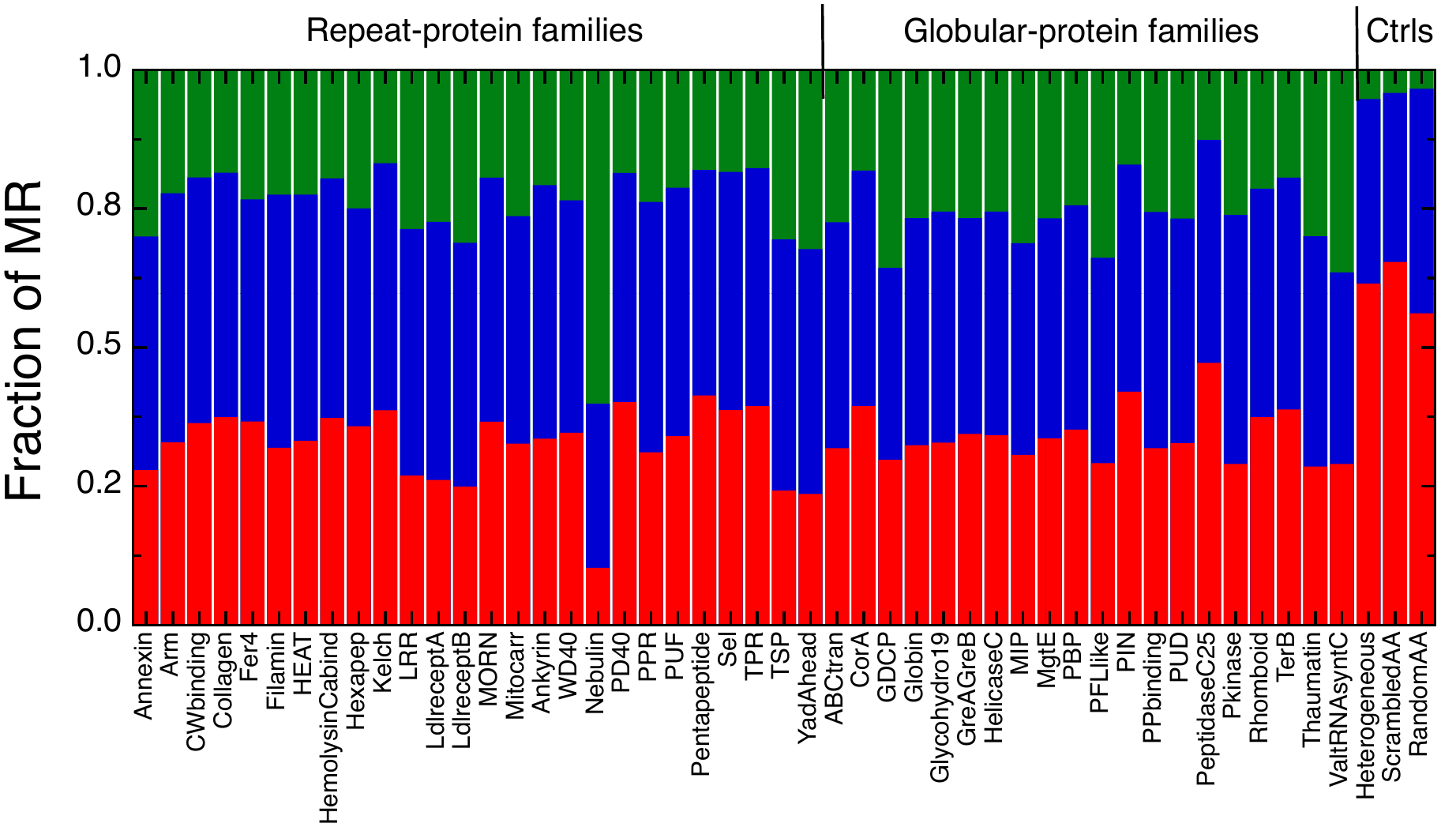}
  \caption{Fraction of MR in each class for all protein families analyzed. Nested maximal repeats (NE, green), Non-Nested maximal repeats (NN, blue) and Super-maximal repeats (SMR, red) were computed in each natural family of either globular or repeat-protein classes. Ctrls indicates the three control groups of artificial families (see main text).} 
  \label{fig:fractionMR_ByFamily}
\end{figure}

In order to analyze the structure of MR patterns in natural protein sequences we concentrate in 46 abundant and curated protein families. Each family contains between 924 and 38,342 non-redundant sequences, and between 805,684 and 23,670,587 amino acids (Table S2), making a grand total of 696,114 strings and 434,447,858 amino acids. We analyze families for which recurrent structural repetitions are annotated (usually called ''repeat-proteins'' \cite{DiDomenico}), and families for which no repetitions were reported (globular proteins). For each family the distribution of MR was calculated and each MR classified as either SMR, NN or NE (see Fig.\ref{fig:MER_Example}). The relative population of each MR class in each family is shown in Fig. \ref{fig:fractionMR_ByFamily}. Overall, the distribution of MR types appears roughly constant between families: most of the MR are non-nested (NN), around 20\% are nested (NE), and about 25\% are true super-maximal repeats (SMR). This holds irrespectively of the common classification of repeat vs globular protein family, indicating that the overall nesting architecture of MR is a general characteristic of all protein sequences. One clear exception is the Nebulin family, for which we identify an overabundance of NE repeats. It was previously reported that the repetitions found in this family can be described as short repeats within longer repeats \cite{Bjorklund:2010kl}, which we capture as nested occurrences (Fig.  \ref{fig:fractionMR_ByFamily}). 

To test whether the distributions of MR types is random or characteristic of natural protein families, we constructed three control groups of sequences: $RandomAA$ is a set of sequences drawn entirely by chance of equally probable 20 letters, $ScrambledAA$ is an exhaustive permutation of the amino acids of one natural family, thus conserving the natural bias in the amino acid composition \cite{Krick:2014oq} and $Heterogeneous$ is a set of natural sequences picked at random from all the families (see Supplementary Methods for details). All three control groups show a common distribution of abundance of MR categories, with SMR being the most prevalent and only a minority of NE (Figure \ref{fig:fractionMR_ByFamily}). For the $RandomAA$ and the $ScrambledAA$ controls it can be expected that most of the MR will not be found in the NE category, as the nesting probability of MR decreases exponentially in random strings \cite{Crochemore}, and thus SMR will prevail. However, the $Heterogeneous$ picking of sequences from various families shows a similar distribution, hinting that the nesting patterns are properties emerging from the grouping of sequences, and are not found at the level of individual proteins.

\begin{figure}
 \center
  \includegraphics[width=0.9\textwidth]{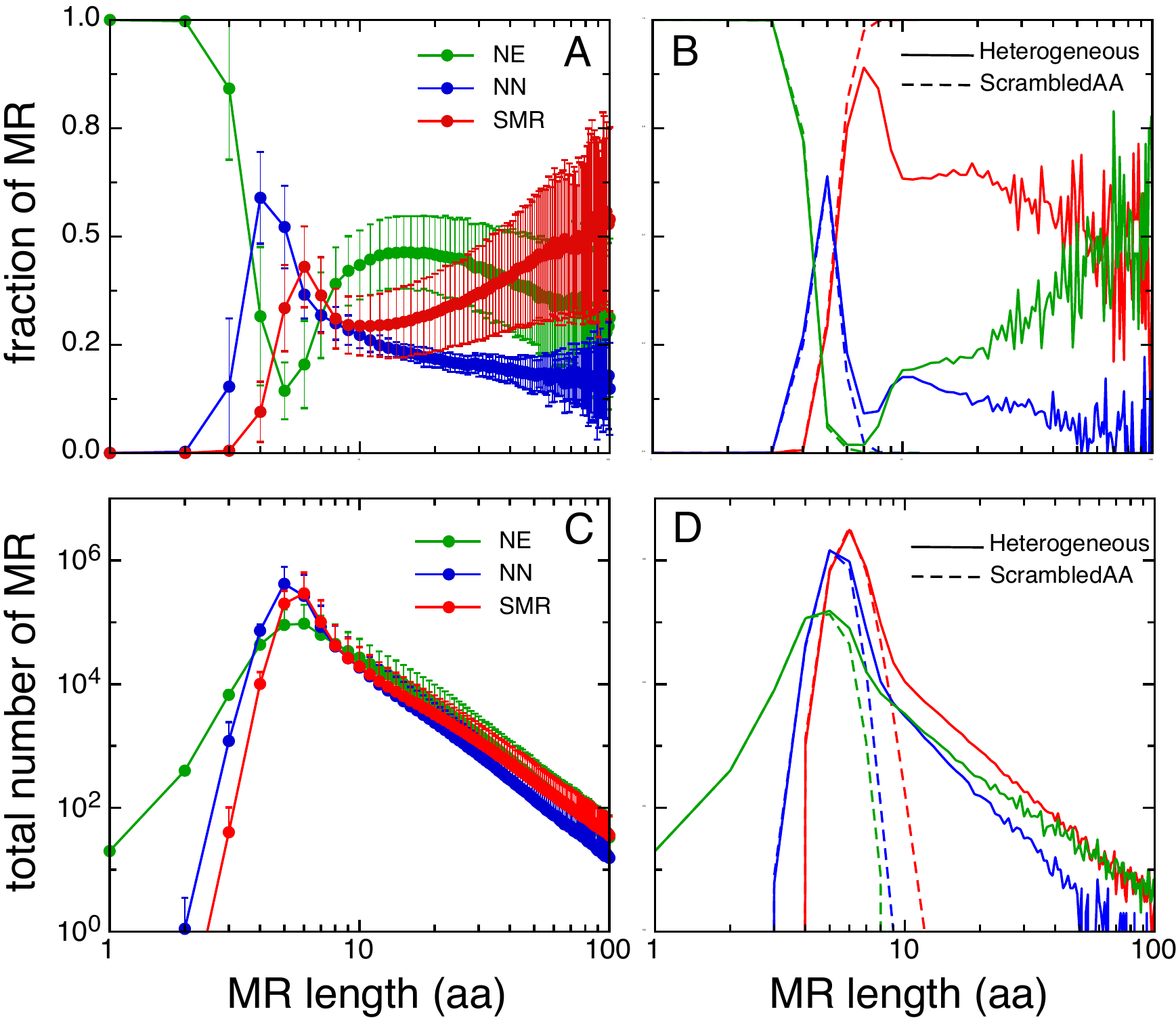}
  \caption{Distribution of MR patterns in each subset. The MRs were calculated for each family and the mean and standard deviation of all families is shown. The relative abundance at different lengths is shown in A) and the total abundance in C). Equivalent calculations of the control families are shown in B) and D). } 
  \label{fig:fractionMR_totalNumber_ByLength}
\end{figure}

The abundance of distinct types of MR in the families could depend on the length of the MR set under scrutiny, as shorter MRs are trivially more prevalent in any string. Figure \ref{fig:fractionMR_totalNumber_ByLength} A shows that all of the repetitions of length 1 and 2 amino acids are nested in longer MR in all families. The fraction of NE drops to about 10\% at length 5 and then grows to about 40\% for lengths of few decades. The non-nested (NN) repeats are most prevalent at length of 4 to 5 amino acids and the SMRs are the most abundant when the longer MRs are considered. The relative abundance of each MR class shows a complicated length dependence, that we find consistently in each family. It is expected that SMR will have to be the most prevalent class at the longest lengths, as every NN or NE is ultimately contained within an a larger SMR (Figure \ref{fig:MER_Example}). The same analysis performed on the random set indeed shows that SMR is the only class of MR at lengths larger than 8 amino acids, with NN being the most prevalent at length 5 and NE absent above length 7 (Fig \ref{fig:fractionMR_totalNumber_ByLength} B, dashed lines). In contrast, when multiple sequences are grouped into an artificial control family, NN and NE persist up to length 100 (Fig \ref{fig:fractionMR_totalNumber_ByLength} B, continuous lines). 

If there is structure in the architecture of the repetitions in a finite string, it is expected that not all MR will be equally abundant. We quantified the total number of different patterns in all the MR classes in all families. As can be seen in Figure \ref{fig:fractionMR_totalNumber_ByLength} C, all of the MR of size 1 (20 single amino acids) are present in all families, and every occurrence is nested in longer MRs. From the tenths of millions of distinct MRs found, most of them have lengths between 4 and 8 amino acids, being the most prevalent SMR and NN types of 4 to 6 amino acids. Notably, for MR longer than 10 residues, the distribution appears to follows a power law where $MR_{abundance} \approx a*length(MR)^\gamma$. The $\gamma$ exponent is about -2.6 regardless of the MR class. This is clearly not the case for the control randomized set, where there are no MRs larger than 12 amino acids and the $\gamma$ exponent is about -10 (Fig \ref{fig:fractionMR_totalNumber_ByLength} D, dashed lines). When multiple sequences are grouped into an artificial control family $\gamma$ is around -3.8 (Fig \ref{fig:fractionMR_totalNumber_ByLength} D, continuous lines). The $\gamma$ exponent appears to be similar for each and every protein family (Fig. S1).

\begin{figure}
 \center
  \includegraphics[width=0.8\textwidth]{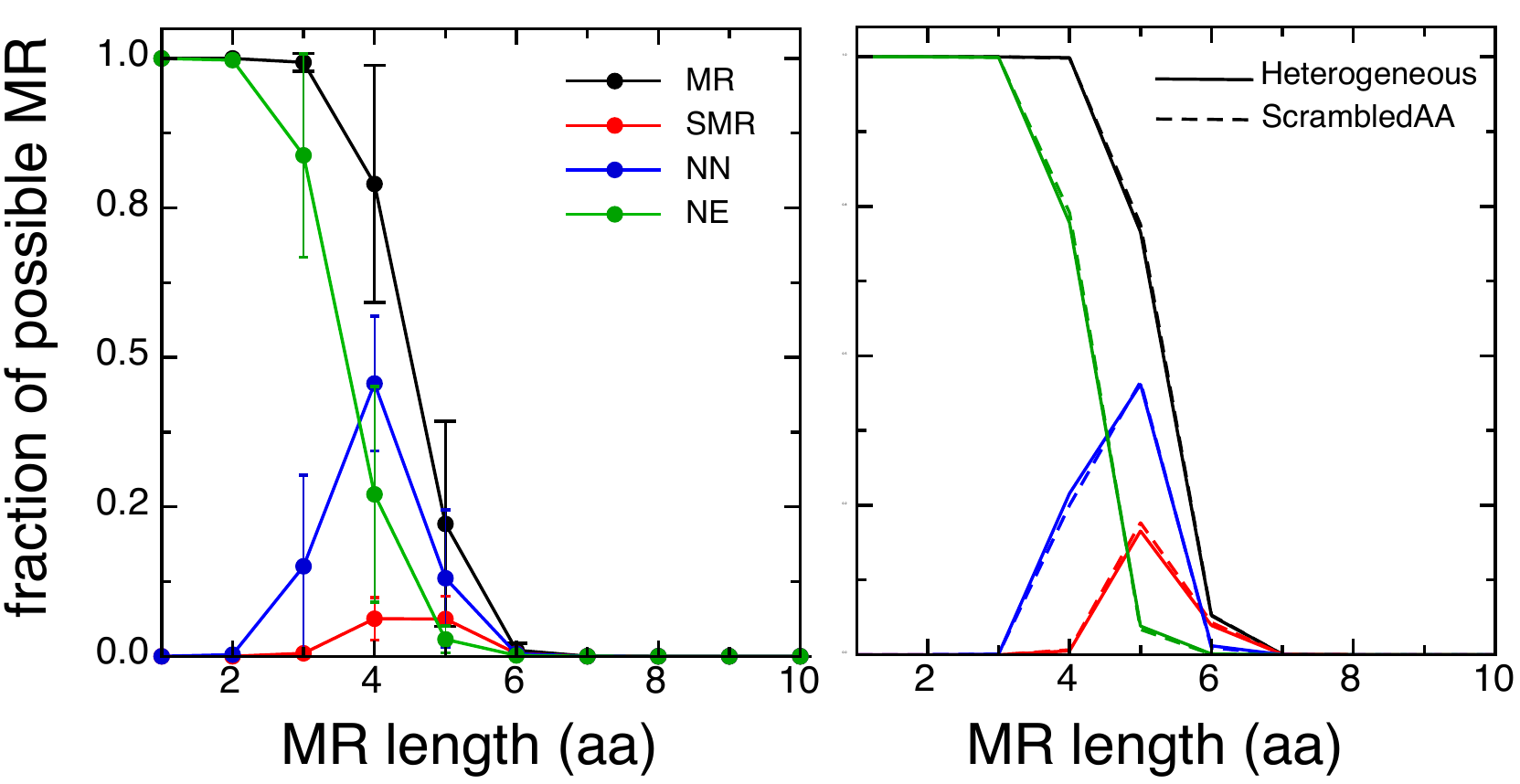}
  \caption{Coverage of the sequence space. The MRs were calculated for each family and the fraction of the total possible patterns is shown for each subset. The mean and standard deviation of all families is shown in A) and for control groups in B)}
  \label{fig:fractionTheoricalMR_ByLength}
\end{figure}

To set the length scale for optimal MR evaluation, we calculated the fraction of the possible strings that are present as MR in natural sequences. All of the single and possible pairs and triplets of amino acids are found in the natural dataset, and these are typically nested in longer MRs (Fig. \ref{fig:fractionTheoricalMR_ByLength} A). As the possible sequences grow as $20^N$, the coverage of the sequence space precipitously drops and only few of the possible MRs of length longer than 6 are found. Both control sets show a slightly higher coverage of the sequence space than the natural families, and SMR and NN are found in larger proportions at length 5 amino acids (Fig. \ref{fig:fractionTheoricalMR_ByLength} B). Notably, both the random and the artificial family sets display equivalent coverage of the possible sequences at short distance, suggesting that natural proteins look effectively random at lengths shorter than 4 amino acids. The artificial grouping of sequences in the Heterogeneous control explores the sequence space equally well as the random set, up to a length of 7 amino acids (Fig. \ref{fig:fractionTheoricalMR_ByLength} B). 

\subsection{Sequence coverage and Familiarity for MR subsets}

In general, all protein families analyzed contain a similar distribution of MR patterns in terms of length and MR class (Fig. \ref{fig:fractionMR_totalNumber_ByLength}). However, the specific sequences of the MR sets of each family can be very different, as they account for an almost insignificant proportion of the possible amino acid strings for lengths larger than 6 residues (Fig. \ref{fig:fractionTheoricalMR_ByLength}). To evaluate how distinct MR sets can account for the occurrence of specific patterns in natural protein sequences, we developed the continuous functions $coverage$ and $familiarity$ \cite{Turjanski2016}. Briefly, the function $coverage:\Sigma^* \times \+P(\Sigma^*) \to \Q$ is defined for any sequence $s$ and any set of sequences $R$

\noindent
\begin{equation}
coverage(s, R) =
\frac{\#\{j:\exists i \in \N , \exists r \in R,\ s[i..i+|r|-1] = r \} }{ |s|}.
\end{equation}

\begin{figure}[h!]
 \center
  \includegraphics[width=0.9\textwidth]{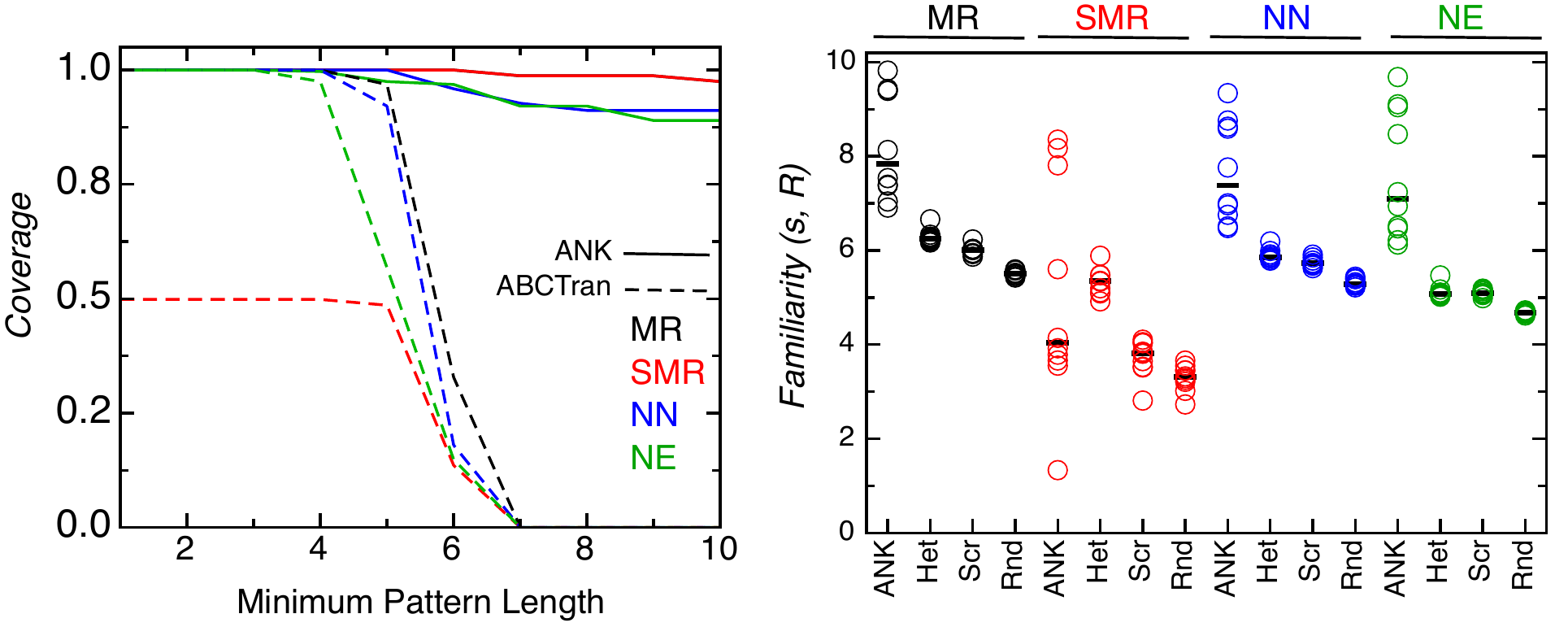}
  \caption{Evaluating $Coverage(s,R)$ and $Familiarity(s,t)$ functions on natural proteins. A) Values for $Coverage$ for the sequence $s$ corresponding to the natural protein I$\kappa$B$\alpha$ of $H. sapiens$, with distinct MR subsets $t$. B) Values for $Familiarity$ for 10 natural protein sequences $s$ annotated to have Ankyrin repeats, evaluated with different MR subsets from the ANK family and three control groups.} 
  \label{fig:Ank_familiarity}
\end{figure}

$coverage(s, R)$ is a rational number between $0$ and $1$. Figure \ref{fig:Ank_familiarity} shows $coverage(s, R)$ evaluated on the string of the natural protein I$\kappa$B$\alpha$ of $H. sapiens$ with distinct MR subsets. The sequence can be covered fully with most short MRs, and is larger than 0.9 for all MR subsets originating from the ANK family up minimum pattern length 10. In contrast, the coverage drops to zero at minimum pattern 7 amino acids when the MR sets are originated from the ABCTran family. This result is not surprising as I$\kappa$B proteins are annotated to contain ankyrin repeat regions (grouped in the ANK family), and none ABCtran family signatures \cite{Trelle:2016vn}. 

The {\em familiarity} function $\Sigma^* \times \Sigma^* \to \Q$ measures how much of a sequence is covered by a set of MRs.  For any sequence $s$ and any sequence $t$,

\begin{equation}
familiarity(s,t)=\frac{coverage(s,\M(t,0))+ coverage(s,\M(t,|s|))}{2} + \sum_{i=1}^{|s|-1} coverage(s,\M(t,i))
\end{equation}

where $\M(t, n)$ denotes the set of MRs from $t$ of lengths greater than or equal to $n$. $\M(t, 0)$, by definition, gives all the blocks of the sequence $t$. $familiarity$ requires the values of $coverage(s,\M(t,i))$ for each  $i$ in $  [0..|s|]$, which we find is enough to limit to $[0..10]$, allowing for a robust comparison of sequences $s$ of different lengths \cite{Turjanski2016}. $familiarity (s,t)$ is thus a rational number between $0$ (when not a single part of $s$ can be covered by MRs of $t$, only possible for disjoint alphabets) and $10$ (when the whole $s$ can be covered with MRs of $t$). As the second argument $t$ of the $familiarity$ function we denote the concatenation of all the sequences of a group separated by the distinguished symbol ($+$) (\textbf{Definition 2}).

Figure \ref{fig:Ank_familiarity}B shows the values for $familiarity$ evaluated over 10 natural test sequences from the Ankyrin family with different sets of MRs. All of these sequences score over 6 when the MR set $t$ originates from the ANK family, as expected, as these sequences are annotated to have ankyrin regions (Table S1). However, the $familiarity$ is around 6 when $t$ are constructed from control sequences. These values of $familiarity$ originate from the common underlying structure of the patterns of both natural and random sequences up to a length of 5 amino acids (\textit{vide supra}). Both NE and NN subsets account for these distinctions and the values spread for the SMR subsets (Fig. \ref{fig:Ank_familiarity}). Thus, these sequences can be similarly well described with the structural patterns of the NE and NN subclasses.

\begin{figure}
 \center
  \includegraphics[width=0.9\textwidth]{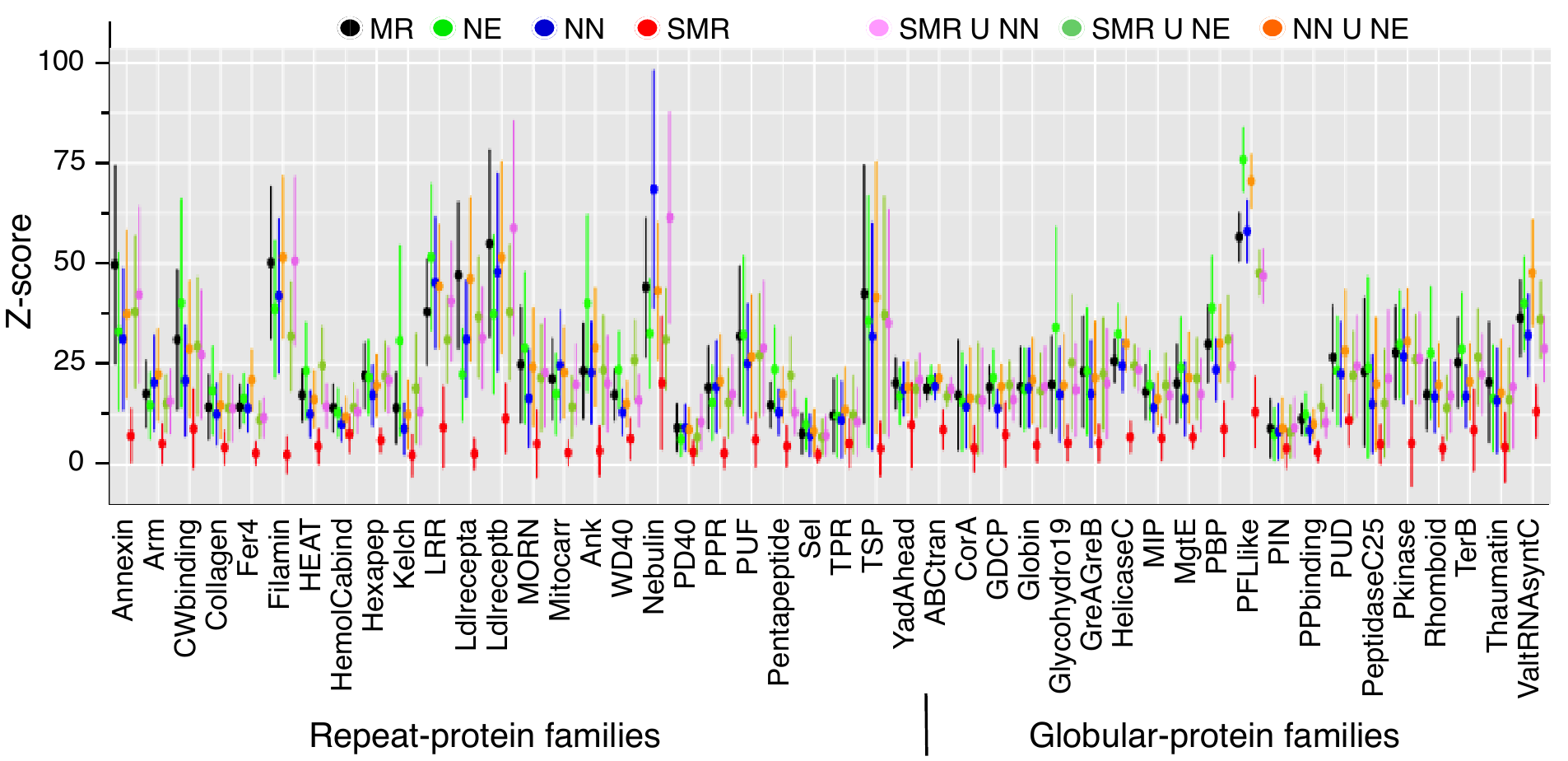}
  \caption{Evaluating $Familiarity(s,t)$ for natural sequences with random MR subsets. $Familiarity(s,t)$ was computed for ten $s$ sequences of each natural protein family and the set $t$ of ScrambledAA control group. The mean and standard deviation of the Zscore distributions are shown, computed with the NN, NE and SMR subsets and all the pair-unions of these SMR$\cup$NN, SMR$\cup$NE, NN$\cup$NE.  } 
  \label{fig:zscore}
\end{figure}

Extant natural protein sequences encode distinct functional \textit{domains}. These are usually reflected as common structural patterns that persist over evolutionary times, and may be sometimes artificially decoupled along the amino acid strings \cite{Espada2015}. These biological lumping must be related with the MR patterns found in the sequence descriptions. To investigate how MR patterns of the families differ from random strings, we computed $familiarity (s,t)$ for 10 test sequences that are annotated to belong to each of the 46 families under scrutiny, but that are not present in $t$. To evaluate the MR subsets on common grounds, we compare the Z-scores of the $familiarity (s,t)$ distribution of test sequences with respect to the random sets (Fig. \ref{fig:zscore}). Both the NN and NE subsets are excellent distinguishing the test proteins from random sequences, as the mean Z-score is above 10 for most families. In most cases, both NE and NN are as good as the whole MR set. Some families show consistently larger Z-scores (Nebulin and PPLlike), with a somehow larger mean values for the families grouped as repeat-proteins (Fig. \ref{fig:zscore}). For all families, the Z-scores of the SMR subsets is lower, indicating that the test sequences cannot be well explained with these patterns. To analyze if combinations of the MR subsets significantly alter the $familiarity (s,t)$ scoring, we constructed all pair-combinations of NN, NE, and SMR in $t$ and found that none of these significantly perturb the results (Fig. \ref{fig:zscore}).

\begin{figure}
 \center
  \includegraphics[width=0.8\textwidth]{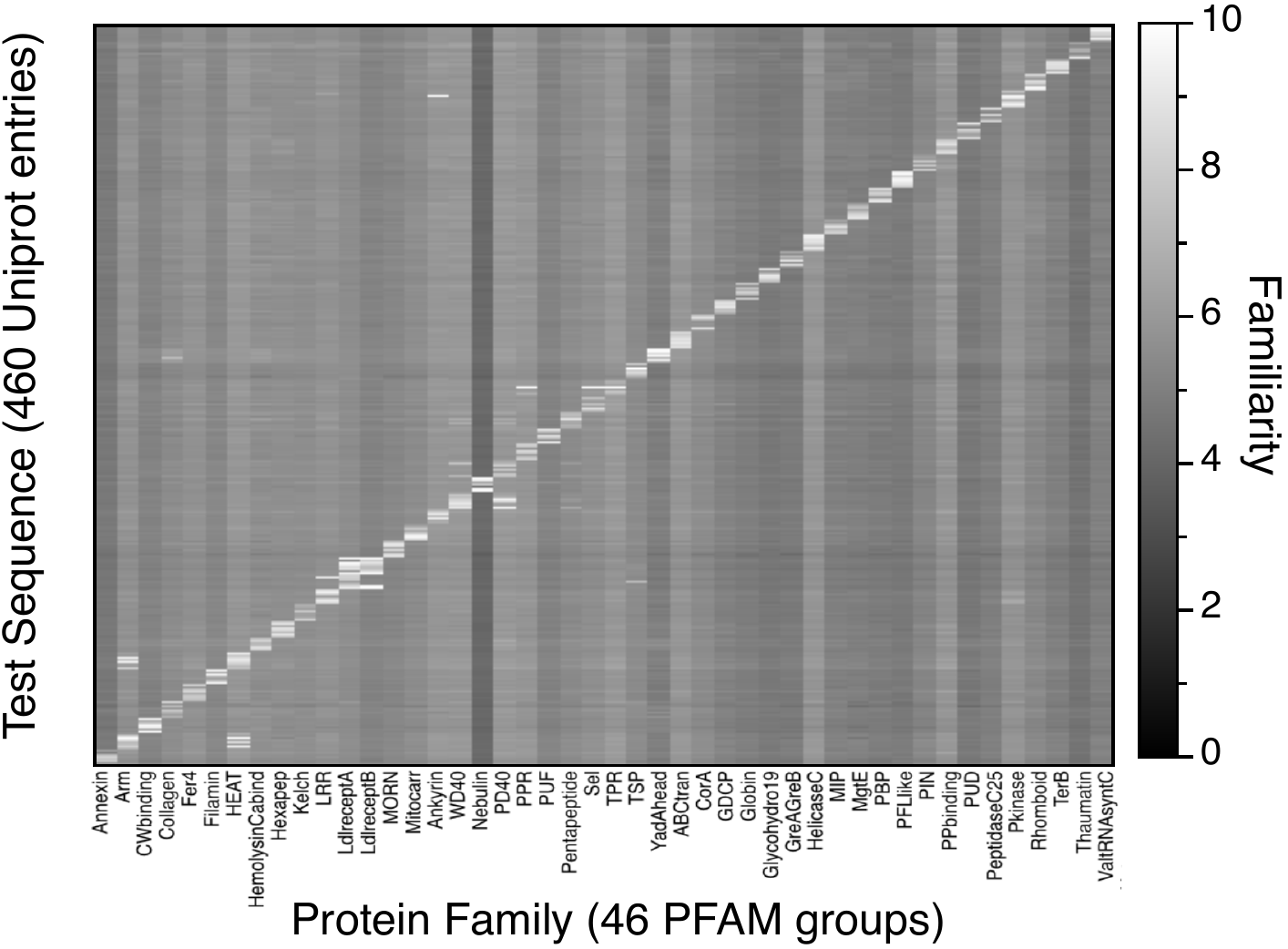}
  \caption{Evaluating $Familiarity(s,t)$ for natural sequences and natural families. $Familiarity (s,t)$ was computed for 460 test sequences $s$ and 46 $t$ families with the SMR$\cup$NN subsets of MR. The unique Uniprot entries are ordered according to the presence of at least one PFAM domain. } 
  \label{fig:matrizall}
\end{figure}

To investigate how MR patterns between the families overlap, we computed $familiarity (s, t)$ for 460 test sequences, 10 $s$ for each of the 46 $t$ families under scrutiny (Table S1). In Fig. \ref{fig:matrizall} the unique Uniprot sequence entries are ordered according to the presence of at least one PFAM domain. The strong diagonal of high $familiarity (s,t)$values thus reflects that the PFAM grouping is consistent with the definitions, the computation and the scorings we propose. Some families display consistently low $familiarity$ towards all sequences (Nebulin), and some consistently higher values (HelicaseC). It is also apparent that some families are clearly related (LdlReceptA and LdlReceptB), even when their historical naming differs (ARM and HEAT). In some cases, a sequence displays significant $familiarity$ to more than one family, hinting to the presence of multiple biological domains. In some other cases, groups of test sequences only display $familiarity$ towards one family (TSP). The multiple ''bands'' that are apparent in the representation of the data in Fig. \ref{fig:matrizall} are probably not random but a manifestation of deeper structure in the original data, which deserves further investigation but is out of the scope of this report. We note that the results are robust to the subsets of MRs that are used to compare the sequences and combinations thereof (Fig S2).

\section*{Discussion}

For \textit{genetic information} to be a meaningful modern concept, natural protein sequences cannot be just random strings of amino acids \cite{10.2307/188717, ADAMI20043, infobio_GodfreySmith}. Spontaneous, fast and robust folding of polypeptide chains is the organic way in which structural patterns emerge from amino acid sequences in certain environments \cite{Ferreiro:2014dq}. The search for underlying \textit{folding codes} to de/construct the folding energy landscapes involves the realization that natural sequences are fundamentally distinguishable from random strings \cite{Bryngelson:1987tg}, yet the actual correspondences are clearly complex, as they involve a myriad of small, non-local, interactions. Effective ways of reverse engineering folding have been achieved at different levels of description with clever heuristics \cite{Munoz:1999hc, Robustelli:2018ij, Leaver-Fay:2011bs}, indicating that it is possible to deconvolute the physical phenomenon without directly recurring to the fundamental quantum mechanical level. Contemplating the historical footprints in the extant sequences has led to useful approximations for the exploration of the energy landscapes of structural \cite{Schafer:2014cr} as well as the sequence spaces \cite{Bornberg-Bauer:1999mi}. 

All known terrestrial proteins can be described as linear repetitions of amino acids. We searched for patterns and groupings of patterns in natural protein sequences using a mathematically rigorous definition for 'repetition' (MR, \textbf{Definition 1}), an efficient algorithmic implementation (Suppl. code2) and a robust scoring system with no adjustable parameters \cite{Turjanski2016}. We propose that the MR set computed for a group of sequences can be well-separated into disjoint classes (Fig. \ref{fig:MER_Example}). Each MR is either super-maximal (SMR), nested (NE) or non-nested (NN) according to the patterns of occurrence in a given set of sequences. The relative population of each MR class in natural protein families is similar and clearly distinguishable from the randomized control groups of sequences (Fig. \ref{fig:fractionMR_ByFamily}). The random grouping of natural sequences in artificial families displays similar total MR fractions as the controls in which the sequences themselves are randomized, indicating that the nesting patterns of the repetitions is the main object underlying the distributions in the natural sets. Indeed, the occurrence of repeats shorter than 5 amino acids is equivalent in the natural and artificial sets (Fig. \ref{fig:fractionMR_totalNumber_ByLength}), covering the sequence space as expected for the exhaustive exploration of patters in random sets of similar, finite, size (Fig. \ref{fig:fractionTheoricalMR_ByLength})\cite{Lavelle:2010uq}. As the sequence space grows exponentially with string length, almost none of it is covered by repetitions larger than 5 amino acids. However, the occurrence of patterns of repetitions in natural sequences is clearly not random in any family and most of the changes in the distributions of MRs occurs between 5 and 10 amino acids (Fig. \ref{fig:fractionMR_totalNumber_ByLength}). Perhaps it is not a coincidence that regular secondary structure elements occur at this length scale \cite{Parra:2013bh}, as well as the critical window at which good structure prediction heuristics work and \textit{foldons} are predicted to emerge \cite{Panchenko:1997fv}. The patterns of repetitions larger than 10 amino acids can be crudely described by a power law distribution in all natural protein families. The proportionality $\gamma$ exponent is about -2.6 for all families and MR subsets (Fig. \ref{fig:fractionMR_totalNumber_ByLength}, Fig. S1). This apparent scale invariant distribution of structure in natural proteins was previously observed at the tertiary level, and related to the alleged fractal geometries of natural folds \cite{PhysRevLett.45.1456, Lewis:1985kl, Reuveni:2010qa, Kornev:2018kx}. Comparable exponents were also reported for the distribution of voids in the interior of protein \textit{swiss-cheese} globules \cite{Chowdary:2009dz}. Thus, there is an apparent common structure of amino acid patterns larger than 10 residues that can be detected in the primary structure of protein families and at the tertiary level of individual folded proteins. 

The search and match of MRs can be efficiently used to characterize the structure of any sequence $s$ with respect to a set of sequences $t$, by computing the $familiarity(s,t)$ function \cite{Turjanski2016}. Both the NE y NN subsets of MRs are as good descriptors of $familiarity(s,t)$ as the whole set (Fig. \ref{fig:zscore}). The overall patterns of repetitions shorter than 10 residues is enough to account for the differences between natural and random sequences by more than 10 standard deviations (Fig. \ref{fig:zscore}). The scoring we put forward is robust to the combinations of MR subsets, and the exhaustive search of the SMR$\cup$NN subset can be implemented with an algorithm whose computational complexity is  $O(n)$ (Suppl. code2). 

Extant natural protein sequences encode \textit{functional domains} of finite size \cite{Espada2015}. The biological accretion of \textit{functional information} can be expected to be detectable at the lengths scales in which proteins differ from the occurrence of patterns in random strings \cite{Dayhoff:1976fk}. Computing $familiarity(s,t)$ for groups of existing proteins indeed reflect exciting patterns of common structure that are discernible at the length scales of 5$\smile$10 amino acids (Fig. \ref{fig:matrizall}). The $familiarity (s,t)$ distributions are robust to the MR subsets used and indicate that PFAM grouping is consistent with the definitions, the computation and the scorings we propose (Fig. S2). In some cases, a sequence displays significant $familiarity$ to more than one family, hinting to the presence of multiple biological domains. In some other cases, groups of test sequences score consistent $familiarity$ towards one family. Presumably evolutionary relationships between groups of sequences can also be detected as groups that score consonant between PFAM families (Fig. \ref{fig:matrizall}). Since $familiarity (s,t)$ is a well-defined continuos function and the MR search can be exhaustively computed with existing computers, it could be used as a general tool to explore the biological relationships between arbitrary groups of sequences. Developing appropriate metrics in the sequence space \cite{doi:10.1093/bib/bbt070} together with efficient search strategies can hint at the length scales in which the \textit{natural coding} of biological information occurs \cite{manoloqv, Kirschner:2000fu, Ferreiro:2018ve, salmon}. 

\acknowledgement

We thank Pablo Rago and Vero Becher for their support and insightful discussions, and Nacho S\'anchez for his persistent challenges. This work was supported by the Consejo Nacional de Investigaciones Cient\'ificas y T\'ecnicas de Argentina (CONICET), the Agencia Nacional de Promoci\'on Cient\'ifica y Tecnol\'ogica (ANPCyT), Ecos-Sud and NAI-Enigma.

\section*{Supporting Information}
Supporting information includes Table S1 and S2 with the protein families detailed data, pseudocode and the algorithmic implementation of MR search and match and figures S1 and S2 of the statistics of MR occurrences.


\subsection*{TOC image}

\medskip
	\begin{figure}[h!]
\centering
	\includegraphics[width=0.3\textwidth]{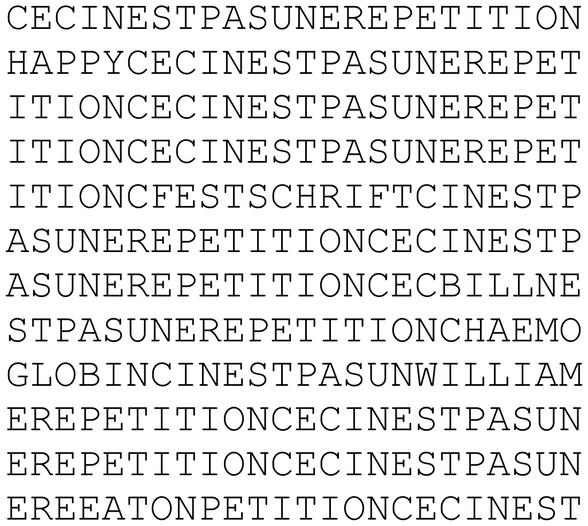}
	\label{fig:toc}
\end{figure}
\medskip

%


\bibliography{Bibliography.bib}

\end{document}